\documentclass[12pt]{iopart}

\usepackage{iopams}  

\usepackage{graphicx}
\usepackage{hyperref}

\def\s{\sigma}
\def\c{\hat{c}^{}}
\def\cc{\hat{c}^{+}}
\def\wn{i\omega_n}
\def\e{$\varepsilon$ }
\def\g{$\gamma$ }
\def\ge{${\gamma\textrm{-}\varepsilon}$ }
\def\ag{${\alpha\textrm{-}\gamma}$ }

\def\etal{\textit{et al.}}

\begin{document}

\title{Structural ${\gamma\textrm{-}\varepsilon}$ phase transition in Fe-Mn alloys from CPA+DMFT approach}

\author{A S Belozerov$^{1,2}$, A I Poteryaev$^{1,3}$, S L Skornyakov$^{1,2}$ and V~I~Anisimov$^{1,2}$}

\address{$^1$ Miheev Institute of Metal Physics, Russian Academy of Sciences, 620137 Yekaterinburg, Russia}
\address{$^2$ Ural Federal University, 620002 Yekaterinburg, Russia}
\address{$^3$ Institute of Quantum Materials Science, 620075 Yekaterinburg, Russia}

\ead{alexander.s.belozerov@gmail.com}

\begin{abstract}

We present a computational scheme for total energy calculations of disordered alloys with strong electronic correlations. It employs the coherent potential approximation combined with the dynamical mean-field theory and allows one to study the structural transformations. The material-specific Hamiltonians in the Wannier function basis are obtained by density functional theory. The proposed computational scheme is applied to study the ${\gamma\textrm{-}\varepsilon}$ structural transition in paramagnetic Fe-Mn alloys for Mn content from 10 to 20 at.~\%. The electronic correlations are found to play a crucial role in this transition. The calculated transition temperature decreases with increasing Mn content and is in a good agreement with experiment. We demonstrate that in contrast to the ${\alpha\textrm{-}\gamma}$ transition in pure iron, the ${\gamma\textrm{-}\varepsilon}$ transition in Fe-Mn alloys is driven by a combination of kinetic and Coulomb energies. The latter is found to be responsible for the decrease of the ${\gamma\textrm{-}\varepsilon}$ transition temperature with Mn content.

\end{abstract}

\pacs{71.15.Mb, 71.20.Be, 71.27.+a}
\maketitle

\section{Introduction}

Despite the rapid development of computational techniques in the last decades,
a first principles investigation of strongly correlated disordered alloys is still very challenging.
Nowadays, substitutional alloys are mainly studied by two different approaches 
(for recent reviews see~\cite{Ruban2008} or \cite{Rowlands2009} and references therein). 
In the first approach, a large supercell with randomly distributed atoms of different types is employed.
%
This strategy is easy to implement and allows one to investigate the properties related to local geometry.
The main drawbacks of this approach are the high computational costs and discrete set of concentrations. 
The second approach involves the mean-field ideology for description of substitutional alloys.
%
This ideology is commonly implemented using the coherent potential approximation~\cite{Soven67} (CPA).
%
%
The main idea of the CPA
is to provide the same physical properties of one component effective medium as an average 
of alloy species, embedded in this effective medium.
At present, the CPA implementation with the so-called diagonal type of disorder
is considered to be the best local approximation for alloys~\cite{Faulkner1982,Ducastelle1991}.
At the same time, it is clear that the CPA cannot be used to study short-range order effects in alloys, and extensions are needed.

In order to describe the properties of actual alloys from first principles,
the density functional theory (DFT) uses both above mentioned approaches,
which suffer from all DFT problems.
%
%
In particular, the paramagnetic state cannot be properly simulated by the nonmagnetic DFT calculations,
and the DFT alone usually fails to reproduce the properties of strongly correlated systems.
For solution of the former problem, the disordered local moment (DLM) method is widely used~\cite{dlm_method}.
In this method the paramagnetic state is modeled by randomly distributed magnetic moments in a supercell 
with a condition of zero net magnetization. 
The later problem is usually solved using the so-called LDA+$U$ method~\cite{Anisimov1997},
where the strong electronic correlations are treated in a static way.
Application of these methods in combination with CPA gives good results~\cite{cpa+u_cpa+dlm,korotin_cpa+u}.
At the same time, the LDA+$U$ works well for insulators with long-range magnetic order, 
while it is less suitable for metallic systems.
Also additional approximations are required to describe finite temperature effects.

The dynamical mean-field theory~\cite{DMFT} (DMFT) was developed two decades ago, and
presently it is regarded as a very powerful tool for the description of strongly correlated systems.
In this theory a lattice problem with many degrees of freedom is replaced by
a correlated atom (or ion) embedded in the energy-dependent effective medium which has to be determined
self-consistently. 
The finite temperature Green's function formalism is employed for the solution of impurity problem.
It allows one to treat properly both the temperature effects and paramagnetic state.
The combination of the DMFT with density functional theory (DFT+DMFT or LDA+DMFT)
resulted in a good description of the spectral properties of strongly correlated paramagnetic compounds~\cite{LDA+DMFT}.
Afterwards, the DFT+DMFT method was successfully applied to study different properties of real materials
(for review see~\cite{Kotliar2006} and \cite{Held2007}).

The CPA and DMFT methods share the effective medium or mean-field interpretation,
and thus they can be easily combined.
The first application of the DMFT to the Anderson-Hubbard model (Hubbard model with disorder)
was done by Jani\v{s} \etal~\cite{CPA+DMFT_method} who investigated thermodynamic properties and
constructed a phase diagram.
Later, many authors studied the magnetic~\cite{Byczyk2005,Byczuk2003},
spectral~\cite{Byczuk2003,Byczuk2004,Lombardo2006}, and 
thermodynamical properties~\cite{Dobrosavljevic1994,Poteryaev2015} of disordered Hubbard model.
Particularly, they found that the metal-insulator transition can occur in a correlated alloy 
at non-integer filling~\cite{Byczuk2004}, and a system can be driven from weakly to strongly
correlated regime by change of disorder strength or concentration~\cite{Byczuk2004,Poteryaev2015}.
In the framework of \textit{ab~initio} calculations, the CPA+DMFT method was used to study the
spectral and magnetic properties of binary alloys~\cite{Minar2005,Sipr2008,Braun2010}
and Heusler alloys~\cite{Chadov2009}.
The influence of disorder and correlation effects on thermopower
in Na$_x$CoO$_2$ was investigated by Wissgott \textit{et al}~\cite{cpa_NaCO}.
%
%

In this paper, we propose a computational scheme 
for total energy calculations of substitutional alloys with strong electronic correlations.
The scheme is implemented within the CPA+DMFT approach
and applied to study the \ge structural transition in Fe-Mn alloys with manganese content from 10 to 20 at.~$\!$\%.
The Fe-Mn alloys exhibit a variety of interesting properties such as
the shape-memory effect~\cite{sme}, Invar and anti-Invar effects~\cite{invar_antiinvar}.
In addition, these alloys at $15-35$ at.~$\!$\% of Mn were found to possess improved strength and ductility
making them the basis for the transformation- and twinning-induced plasticity (TRIP and TWIP) steels~\cite{twip_and_trip}.
%
%
Starting from 10~at.~\% of Mn, upon cooling the $\gamma$~phase with the face-centered cubic lattice~(fcc) 
transforms martensitically to the $\varepsilon$~phase with the hexagonal close packed~(hcp) lattice~\cite{FeMn_phase_diagram}.
%
%
Up to 23~at.~\% of Mn, this transition occurs in the paramagnetic region~\cite{Cotes2004},
which significantly complicates the use of most electronic structure calculation methods.
%

The previous first principles studies of Fe-Mn alloys were performed using the CPA approach implemented 
within EMTO formalism~\cite{Music2007,Gebhardt2010,Gebhardt2011_1,Gebhardt2011_2,Ekholm2011}
and supercell approach~\cite{Gebhardt2011_1,Ekholm2011,Dick2009,Reyes-Huamantinco2012,Lintzen2013}.
The paramagnetic state was simulated by means of the DLM model.
In these studies, the elastic properties~\cite{Music2007}, magnetic properties~\cite{Ekholm2011}, lattice stability~\cite{Gebhardt2010},
and stacking fault energy~\cite{Dick2009,Reyes-Huamantinco2012} were investigated.
The enthalpies of formation at 0~K were calculated in~\cite{Lintzen2013}.
The influence of Al and Si additions on the elastic properties and lattice stability was investigated
in~\cite{Gebhardt2011_2} and \cite{Gebhardt2011_1}, respectively.
%
%
In all previous studies of Fe-Mn alloys the Coulomb correlations were considered in some average sense within DFT,
while they were demonstrated to play a crucial role in pure iron~\cite{Leonov_bccfcc,Fe_Leonov,Fe_correlations}.

The paper is organized as follows.
In section~\ref{sec:method} we present
a computational scheme of the CPA+DMFT method for real alloys implemented for total energy calculations.
In section~\ref{sec:results} we employ this 
technique to study the phase stability and magnetic properties of Fe-Mn alloys.
Finally, conclusions are presented in section~\ref{sec:conclusions}.

\section{Method\label{sec:method}}

Let us consider a binary alloy A$_{1-x}$B$_x$ with substitutional type of disorder.
It can be described by the Anderson-Hubbard Hamiltonian  
\begin{eqnarray} \label{ham_cpa+dmft}
 \hat{H}  &=&  -\sum_{\langle i,j \rangle} \sum_{\{m\},\s}
       t_{mm'} ( \cc_{im\sigma} \c_{jm'\sigma} + {\rm H.c.}) \nonumber \\
  &+&  \sum_{i,m,\sigma}(\epsilon^i_m-\mu)\, \hat n^i_{m\sigma} 
   + \hat H_{\rm Coul}, 
 \end{eqnarray}
where $\cc_{im\s}$ ($\c_{im\s}$) is the creation (annihilation) operator
of an electron with spin~$\s$ at orbital~$m$ of site~$i$,
${\hat n^i_{m\sigma}=\cc_{im\s} \c_{im\s}}$,
$\mu$ is the chemical potential,
$t_{mm'}$ is the hopping amplitude,
$\epsilon^i_m$ is the on-site energy,
${\rm H.c.}$ denotes the hermitian conjugate of the preceding term.
The last term in Hamiltonian~(\ref{ham_cpa+dmft}) corresponds to the on-site Coulomb interaction,
which is considered in the density-density form:
\begin{eqnarray} \label{Hcoul}
   \hat H_{\rm Coul} = \frac{1}{2} \sum_{i,m,m',\sigma,\sigma'}
   U^i_{mm'\sigma\sigma'} \hat n^i_{m\sigma} \hat n^i_{m'\sigma'},
\end{eqnarray}
where $U^i_{mm'\sigma\sigma'}$ is an element of the Coulomb interaction matrix.
%
%
The on-site potential $\epsilon^i_m$ and Coulomb matrix $U^i_{mm'\sigma\sigma'}$
depend on the site index $i$ and are different for different atomic species.
At the same time, each site can be occupied by atom of type~A with probability 
${(1-x)}$ or of type~B with probability~$x$. 
The hopping amplitudes are assumed to be site-independent,
which implies similar shapes of band structures for constituents A and B. 
This approximation is reasonable for constituents with similar electronic structures,
when the on-site local potentials are close in energy relative to the bandwidth~\cite{Faulkner1982,Ducastelle1991,Abrikosov1998}.

For material specific calculations all parameters of Hamiltonian~(\ref{ham_cpa+dmft})
are to be determined, and we follow the conventional LDA+DMFT prescription~\cite{Anisimov2005,Lechermann2006}.
In this case, the Hamiltonian can be rewritten as
\begin{eqnarray} \label{ham_dft}
 \hat{H} =  \hat H_{\rm DFT}  +  \sum_{i,m,\sigma}(\epsilon^i_m-\mu) \hat n^i_{m\sigma} 
          + \hat H_{\rm Coul} -  \hat H_{\rm DC}.
 \end{eqnarray}
Here, the kinetic contribution is replaced by the DFT Hamiltonian $\hat H_{\rm DFT}$
calculated in a basis of Wannier functions or other localized orbitals.
The on-site local potential $\epsilon^i_m$ can also be found from DFT results
as a center of gravity of orbital $m$ at site $i$
in a supercell calculation.
The following disorder parameter can be introduced as a difference between centers of gravity
for different atomic species:
\begin{eqnarray} \label{dV}
  V_m = \epsilon^{\rm B}_m - \epsilon^{\rm A}_m.
\end{eqnarray}
The constrained DFT method~\cite{Aryasetiawan2006} can be used to determine elements of 
the screened Coulomb interaction matrix $U^i_{mm'\sigma\sigma'}$.
The last term in equation~(\ref{ham_dft}), $\hat H_{\rm DC}$, is introduced
to avoid double counting of the Coulomb interaction already present in $\hat H_{\rm DFT}$. 
The fully localized limit is taken, and
\begin{equation} \label{eq:dc}
  \hat H_{\rm DC} = \sum_i \bar{U}^i ( \hat n^i - \frac{1}{2} ) \equiv \sum_i \epsilon^i_{\rm DC},
\end{equation}
where $\bar{U}^i$ is the average Coulomb interaction, and
$\hat n^i = \sum_{m\sigma} \hat n^i_{m\sigma}$.

Within the CPA+DMFT approach a real alloy is replaced by an effective medium with local Green function
\begin{eqnarray} \label{gloc}
 G&_{\rm med}&(i\omega_n) = \frac{1}{V_{\rm BZ}} 
      \int G_{\rm med}(\textbf{k},i\omega_n)  {\rm d}\textbf{k}      \nonumber                 \\
  & = &\frac{1}{V_{\rm BZ}} 
      \int\frac{{\rm d}\textbf{k}}{ (\mu+i\omega_n) I - H_{\rm DFT}(\textbf{k}) - \Sigma(i\omega_n)},
\end{eqnarray}
where $\omega_n = (2n+1)\pi/\beta$ are the fermionic Matsubara frequencies, 
$\beta$ denote the inverse temperature, 
$I$ is the unit matrix, 
$\Sigma(i\omega_n)$ is the local effective potential or self-energy,
which has to be determined self-consistently.
The integration is performed over the first Brillouin zone of volume~$V_{\rm BZ}$.
%
%
In contrast to the conventional CPA approach,
the self-energy now contains information not only about disorder,
but also about electronic correlations.
Using the Dyson equation one can obtain the bath Green function
\begin{eqnarray} \label{dyson1}
  \mathcal{G}_0^{-1}(i\omega_n) = G_{\rm med}^{-1}(i\omega_n) + \Sigma(i\omega_n),
\end{eqnarray}
%
%
which is required to calculate the impurity Green functions $G_{\rm A}(i\omega_n)$ and $G_{\rm B}(i\omega_n)$.
The action of impurity embedded in the effective medium is
%
\begin{eqnarray} \label{eq:action}   \nonumber
  S_i & =\!\! & \hspace{-0.01cm} -\!\! \sum_{nm\s} \! \cc_{m\s}(\wn) [ \wn \!-\! \epsilon_m^i \!-\! \epsilon^i_{\rm DC}\! -\! \Delta_m(\wn) ] \c_{m\s}(\wn)  \\ 
      &   & + \int_0^{\beta} d\tau\, U^i_{mm'\s\s'} \hat n^i_{m\s}(\tau) \hat n^i_{m'\s'}(\tau),
\end{eqnarray}
where $\Delta_m(\wn) = \wn - \mathcal{G}_{0,m}^{-1}(\wn)$ is the hybridization function.
The corresponding impurity Green function can be expressed as
\begin{equation}
  G_i = \frac{ \int_0^{\beta} \c \cc e^{-S_i} \mathcal{D}\c\mathcal{D}\cc }
             { \int_0^{\beta} e^{-S_i} \mathcal{D}\c\mathcal{D}\cc }.
  \label{eq:G_a}
\end{equation}
According to the CPA ideology, the local Green function of effective medium is interpreted 
as a weighted sum of impurity Green functions:
\begin{eqnarray} \label{gf33}
  G_{\rm med}(i\omega_n) = (1-x)G_{\rm A}(i\omega_n) + x\, G_{\rm B}(i\omega_n).
\end{eqnarray}
%
%
Having obtained $G_{\rm med}(i\omega_n)$ by equation~(\ref{gf33}),
one can easily compute the new self-energy from the Dyson equation:
%
%
\begin{eqnarray} \label{dyson2}
  \Sigma(i\omega_n) = \mathcal{G}_0^{-1}(i\omega_n) - G_{\rm med}^{-1}(i\omega_n).
\end{eqnarray}
This new effective potential is then used in equation~(\ref{gloc}) to calculate the local Green function of effective 
medium. 
The above equations are iteratively solved until the convergence with respect to the self-energy is achieved.

In the orbital space, the above Green functions, self-energies and other quantities are matrices
of the same size as $\hat H_{\rm DFT}$.
At the same time, they have a block diagonal structure in the orbital space, and hence
solution for different types of orbitals can be performed separately.
For the uncorrelated subspace (Wannier functions of $sp$~character), 
the impurity action in equation~(\ref{eq:action}) becomes Gaussian, and
the impurity Green function can be evaluated straightforwardly:
\begin{eqnarray} 
  G_{i,m}^{-1}(\wn) & = & \mathcal{G}_{0,m}^{-1}(i\omega_n) - \epsilon^i_m  \nonumber  \\
                    & = & \wn - \epsilon^i_m - \Delta_m(\wn).
\end{eqnarray}
To calculate the impurity Green functions in the correlated subspace (the Wannier functions of $d$ character),
the equation~(\ref{eq:G_a}) is to be solved.
The continuous-time quantum Monte-Carlo method~\cite{CT-QMC} was used for the above purpose.

The calculation of total energy has been thoroughly discussed for the LDA+DMFT method~\cite{total_energy}
and for a single band model of disordered system~\cite{Poteryaev2015}.
Following the same way, the total energy in the CPA+DMFT method can be defined as
\begin{eqnarray} \label{tot_en}
E_{\rm total} & = & E_{\rm DFT} + E_{\rm kin}^{\rm CPA+DMFT} + E_{\rm Coul}^{\rm CPA+DMFT} \nonumber \\
               & - & E_{\rm kin}^{0} - E_{\rm DC}.
\end{eqnarray}
Here, the first term is the total energy obtained in self-consistent DFT calculations.
The second term is the CPA+DMFT kinetic energy which can be defined as
\begin{eqnarray} \label{ekin_cpadmft}
E_{\rm kin}^{\rm CPA+DMFT} \! && =
T\sum_{\textbf{k},n} {\rm Tr}[H_{\rm DFT}(\textbf{k}) G_{\rm med}(\textbf{k},i\omega_n)]e^{i\omega_n0^+} \nonumber \\
                                 && + (1-x) \sum_{m\s} \epsilon^{\rm A}_m n^{\rm A}_{m\s} + 
                                          x \sum_{m\s} \epsilon^{\rm B}_m n^{\rm B}_{m\s}, 
\end{eqnarray}
where the first term depends on the effective medium Green function,
which includes disorder and correlation effects;
the last two terms in equation~(\ref{ekin_cpadmft}) represent a contribution to the kinetic energy due to disorder.
The third term in equation~(\ref{tot_en}) corresponds to the Coulomb energy in CPA+DMFT
and can be expressed via double occupancies:
\begin{eqnarray}
E_{\rm Coul}^{\rm CPA+DMFT} =
  (1&& - x) \!\!\sum_{mm'\sigma\sigma'}  U_{mm'\sigma\sigma'}^{\rm A} \langle \hat n_{m\sigma}^{\rm A} \hat n_{m'\sigma'}^{\rm A}\rangle \nonumber \\
  && + x  \sum_{mm'\sigma\sigma'}  U_{mm'\sigma\sigma'}^{\rm B} \langle \hat n_{m\sigma}^{\rm B} \hat n_{m'\sigma'}^{\rm B}\rangle.
\end{eqnarray}
The fourth term on the right-hand side of equation~(\ref{tot_en}) is the sum of DFT
valence-state eigenvalues which is evaluated as the thermal
average of DFT Hamiltonian with the noninteracting DFT Green function:
\begin{eqnarray}
E_{\rm kin}^{0}=T \sum_{\textbf{k},n}  {\rm Tr}[H_{\rm DFT}(\textbf{k}) G^{0}_{\rm med}(\textbf{k},i\omega_n)]e^{i\omega_n0^+},
\end{eqnarray}
where
\begin{eqnarray}
G^{0}_{\rm med}(\textbf{k},i\omega_n) = [(\mu+i\omega_n) I - H_{\rm DFT}(\textbf{k})]^{-1}.
\end{eqnarray}
The last term in equation~(\ref{tot_en}) corresponds to the double-counting energy which
can be written in the fully localized limit as
\begin{eqnarray}
 E_{\rm DC} = (1-x)\frac{\bar{U}^{\rm A}n^{\rm A}(n^{\rm A}-1)}{2} + x\frac{\bar{U}^{\rm B}n^{\rm B}(n^{\rm B}-1)}{2}.
\end{eqnarray}

It should be noted, that the above described CPA+DMFT scheme as well as the expression for total energy
behave correctly in different limiting cases.
Namely, if there is one type of atomic species (${x=0}$ or ${x=1}$)
or atomic species are identical (${V_m=0}$ and ${U^{\rm A}=U^{\rm B}}$)
all equations reduce to the conventional LDA+DMFT ones.
%
%
In the non-interacting limit (${U^{\rm A}=U^{\rm B}=0}$), 
the equations transform to those of classical CPA.

As discussed in the Introduction section, the DMFT together with CPA have already been used to study single-band models and real alloys.
In contrast to studies~\cite{Minar2005,Sipr2008,Braun2010,Chadov2009} where the CPA+DMFT approach was implemented within the Korringa-Kohn-Rostoker method,
in our computational scheme we first calculate material specific parameters for Hamiltonian (1) and then solve it using CPA+DMFT set of equations.
Our scheme is similar to that used in~\cite{cpa_NaCO} where the on-site potential was introduced to mimic the Na potential in Na$_x$CoO$_2$.


\section{Results and discussion\label{sec:results}}

To perform calculations within DFT, we employed 
the full-potential linearized augmented-plane wave method
implemented in the Exciting-plus code (a fork of ELK code with Wannier function projection procedure~\cite{elk}). 
The exchange-correlation potential was considered in the
Perdew-Burke-Ernzerhof form~\cite{PBE} of the generalized gradient approximations (GGA).
The calculations were carried out with the experimental lattice constants ${a_{\rm fcc}= 3.5812}$~\AA\, for the \g phase;
${a_{\rm hcp} = 2.5273}$~\AA\ and ${c_{\rm hcp} = 4.0857}$~\AA\, for the \e phase~\cite{lattice_constants}.
The total energy convergence threshold of $10^{-6}$~Ry was used.
Integration in the reciprocal space was performed using 18$\times$18$\times$18
and 16$\times$16$\times$10\, $\textbf{k}$-point meshes for the \g and \e phases, respectively.
%
%
In nonmagnetic calculations the ground state of $\varepsilon$-Fe is 99 meV/at lower in energy
than the ground state of $\gamma$-Fe.
The supercells with 8 atoms were constructed by doubling all primitive vectors for fcc structure and two vectors in hexagonal plane for hcp structure.
In each supercell, one atom of Fe was substituted by Mn atom.
%
%
The distances between Mn and its nearest periodic image are equal to 5.065~\AA\ and 4.086~\AA\ for the fcc and hcp structures, respectively.
We note that the local relaxation effects are neglected within our scheme. In the case of Fe-Mn alloy, they are expected to be insignificant, since Fe and Mn are neighbours in the periodic table and have close atomic radii (1.26~\AA\ and 1.27~\AA, respectively). A detailed comparison of CPA results with those obtained using large supercells can be found in~\cite{Zhang2010}.

\begin{figure}[t]
\centering
\includegraphics[clip=true, width=0.6\textwidth]{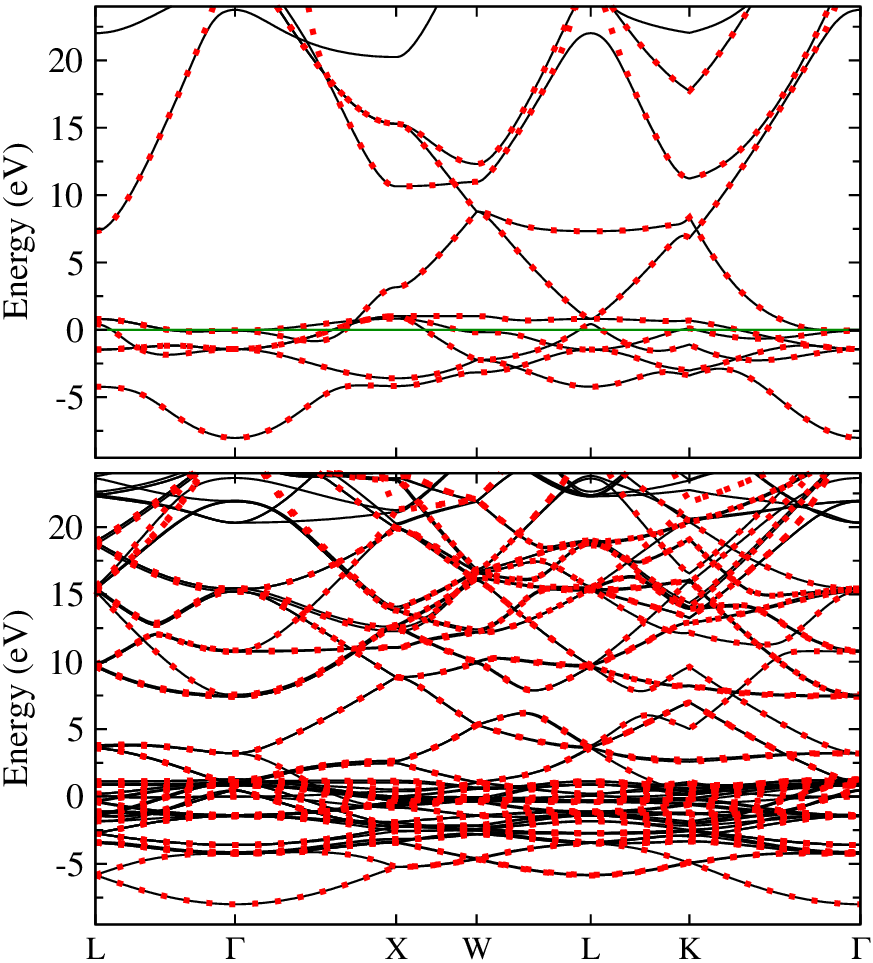}
\caption{ 
  Band structures for fcc~Fe (upper panel) and fcc supercell containing 7 Fe and 1 Mn atoms (lower panel)
  obtained in nonmagnetic GGA calculations (solid lines)
  in comparison with bands corresponding to the constructed
  effective Hamiltonians in the $spd$ Wannier functions basis (dotes).
  The Fermi level is at zero energy.
\label{fig:bands}}
\end{figure}

For CPA+DMFT calculations a localized basis is required, and to this aim,
effective Hamiltonians were constructed for each phase in the basis of Wannier functions.
From converged plane wave data the Wannier functions were built as a projection
of the original Kohn-Sham states to site-centered localized functions of $spd$
character as described in~\cite{Korotin08}.
We note that the obtained Wannier functions are not maximally localized,
that is in fact not necessary for the calculations.
Figure~\ref{fig:bands} shows the original band structure (black lines) for pure fcc Fe (top panel) and 
fcc supercell structure (bottom panel) 
in comparison with bands corresponding to the constructed Wannier Hamiltonians (red dotes).
The Wannier function basis describes well the DFT energy bands up to 18~eV above the Fermi level.

In the following, we use the Wannier function Hamiltonians of pure iron for CPA+DMFT calculations and
refer to them as hosts of corresponding crystal structures. The manganese ions are cited as impurities. 
In this terminology, $\hat H_{\rm DFT}$ from equation~(\ref{ham_dft}) is the host Hamiltonian, and 
$\epsilon^{\rm Fe}$ is already included in $\hat H_{\rm DFT}$. The disorder parameter $V_m$
can be found as a difference between centers of gravity for densities of states (DOS)
for Mn and the most distant Fe atom in supercell calculations.

\begin{table}[b]
    \caption{Disorder parameters for different orbital symmetries obtained in supercell calculations.
    \label{tab:model}}
      \begin{indented}
  \item[] \begin{tabular}{cccc}
\br
      Phase  & $V_s$ (eV)  &  $V_p$ (eV) &  $V_d$ (eV)    \\
\mr
        fcc  &  0.098    &  0.131  &  0.318 \\
        hcp  &  0.151    &  0.161  &  0.362 \\
\br
    \end{tabular}
  \end{indented}
\end{table}

\begin{figure*}
\centering
\includegraphics[clip=true, width=1.1\textwidth]{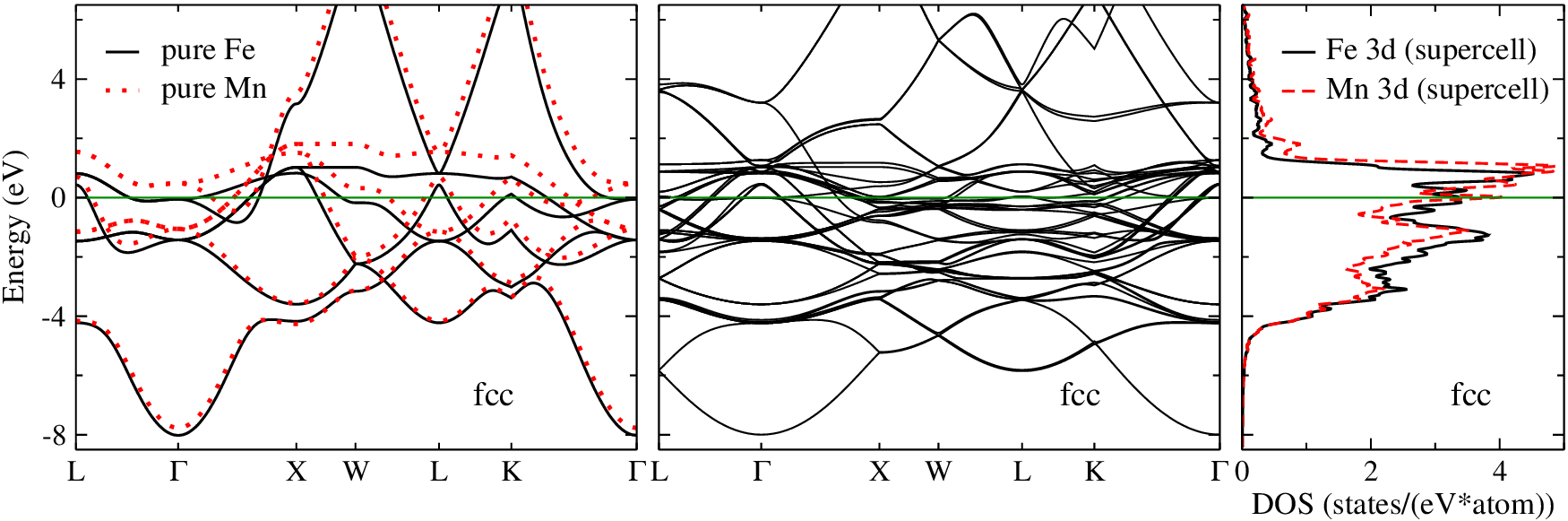}
\includegraphics[clip=true, width=1.1\textwidth]{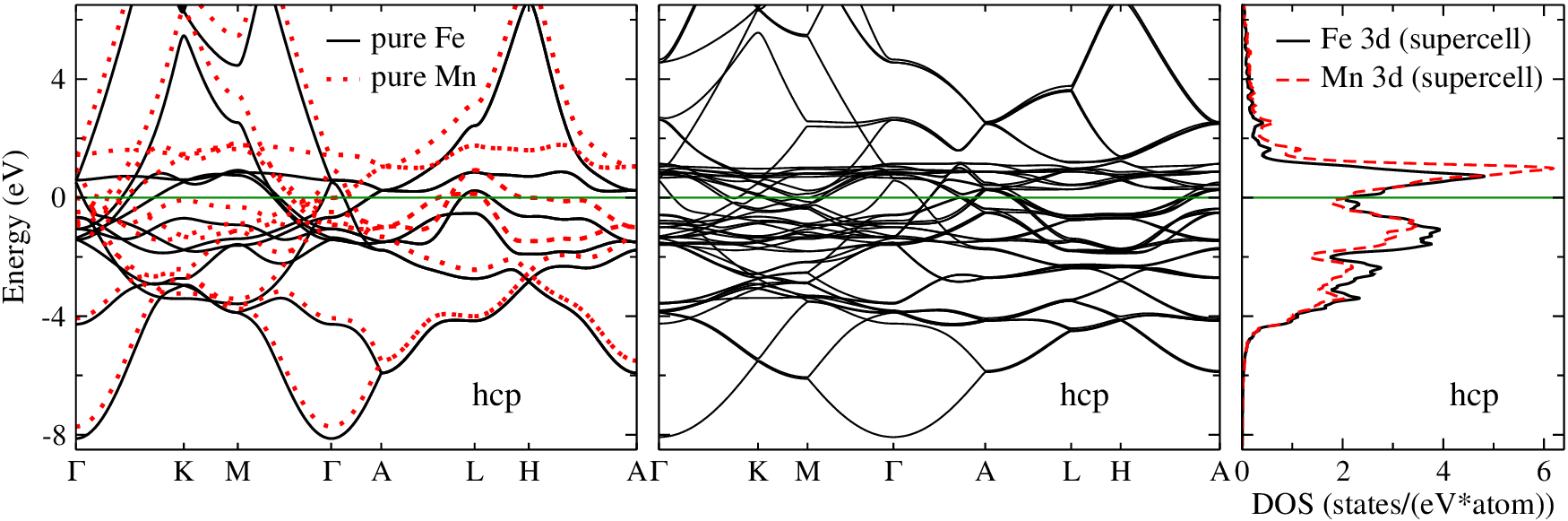}
\caption{ 
  Band structures for pure Fe and Mn (left panels),
  and supercells containing 7 Fe and 1 Mn atoms (central panels)
  obtained in nonmagnetic GGA calculations. 
  Local densities of states for Mn atom and the most distant Fe atom
  in the supercell calculations are presented in the right panels.
  The fcc and hcp structures correspond to the upper and lower panels, respectively.
  The $spd$ Wannier functions basis is employed. 
  The Fermi level is at zero energy.
\label{fig:bands+dos}}
\end{figure*}

Band structures for pure Fe and Mn in fcc and hcp crystal structures
are presented in figure~\ref{fig:bands+dos} (left panels).
One can see that the energy bands of Mn can be described as those of Fe shifted by a constant value.
This value can be regarded as an upper limit for disorder parameter and is about 0.5~eV for $d$ states at $\Gamma$ point.
%
%
%
Comparing the band structures of pure elements and constructed supercells (central panels of figure~\ref{fig:bands+dos}),
one can note a shrinking of the Mn $3d$ bandwidth in the supercells with respect to pure Mn.
%
%
This is supported by the local densities of states (figure~\ref{fig:bands+dos}, right panels) 
for Mn and the most distant Fe atom in the supercells.
The disorder parameters evaluated for different orbitals as the difference between centers of gravity of 
corresponding DOSes are presented in table~\ref{tab:model}.
Since the $s$ and $p$ bands extend beyond the region where they 
are well described by the constructed Wannier functions (see figure~\ref{fig:bands}),
the centers of gravity were calculated using energy window of 15~eV.
Using a wider energy window affects only the values of disorder parameters for $s$ and $p$ states. 
However, as will be discussed further, the disorder parameters $V_s$ and $V_p$ have little influence on the results.

For CPA+DMFT calculations we used the AMULET code~\cite{amulet_code} developed in our group.
Our calculations were carried out with ${U=4}$~eV and ${J=0.9}$~eV obtained by
the constrained density functional theory (cDFT) calculations in the basis of $spd$ Wannier functions~\cite{Belozerov_u_paper}.
This ${U=4}$~eV is in agreement with $U$ from 3 to 4.5~eV 
obtained by constrained random-phase approximation~\cite{cRPA} (cRPA).
The Coulomb interaction within CPA+DMFT was considered in the density-density form and had an atomic structure for $d$ shell.
The Coulomb interaction matrix was parametrized~\cite{Anisimov1997} via Slater integrals $F^0$, $F^2$ and $F^4$
linked to the Hubbard parameter ${U\equiv F^0}$ and Hund's rule coupling ${J\equiv (F^2+F^4)/14}$ with ${F^4/F^2=0.625}$.
Fixed values of $d$ states occupations were used for double-counting terms (see equation~(\ref{eq:dc})).
These values are $n_d^{\rm Fe}$=6.79 (6.84) and $n_d^{\rm Mn}$=5.74 (5.80) for the $\gamma$ ($\epsilon$) phase.
To solve the impurity problems, we employed the hybridization expansion continuous-time 
quantum Monte Carlo (CT-QMC) method~\cite{CT-QMC}.
%

\begin{figure}[t!]
\centering
\includegraphics[clip=true, width=0.56\textwidth]{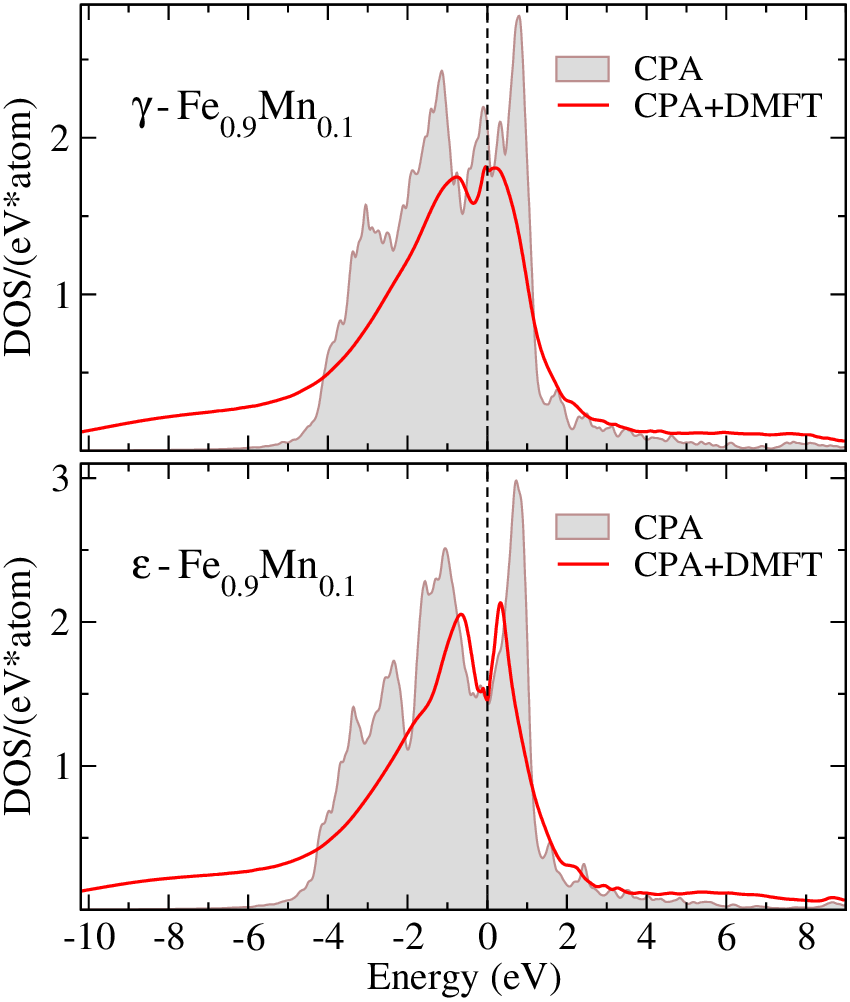}
\caption{ 
Density of $3d$ states obtained by CPA and CPA+DMFT calculations for \g (upper panel) and \e (lower panel) phases of Fe$_{0.9}$Mn$_{0.1}$
at ${\beta=16}$~eV$^{-1}$. The Fermi level is at zero energy.
\label{fig:dos_cpa_dmft}}
\end{figure}

In figure~\ref{fig:dos_cpa_dmft} we present densities of $3d$ states obtained by CPA and CPA+DMFT methods. 
They were calculated as a weighted sum of local densities of states for constituents,
obtained using the Pad\'{e} approximants.
%
Taking into account the electronic correlations by DMFT resulted in a transfer of spectral
weight from the states near the Fermi level to higher energies.
%
%
As in pure bcc iron~\cite{Belozerov_u_paper},
the Hubbard bands are not clearly distinguished since ${U=4}$~eV is less than the bandwidth of about 6~eV.
One can note the similar impact of electronic correlations on density of states as for systems without structural disorder.

\begin{figure}[t!]
\centering
\includegraphics[clip=true, width=0.56\textwidth]{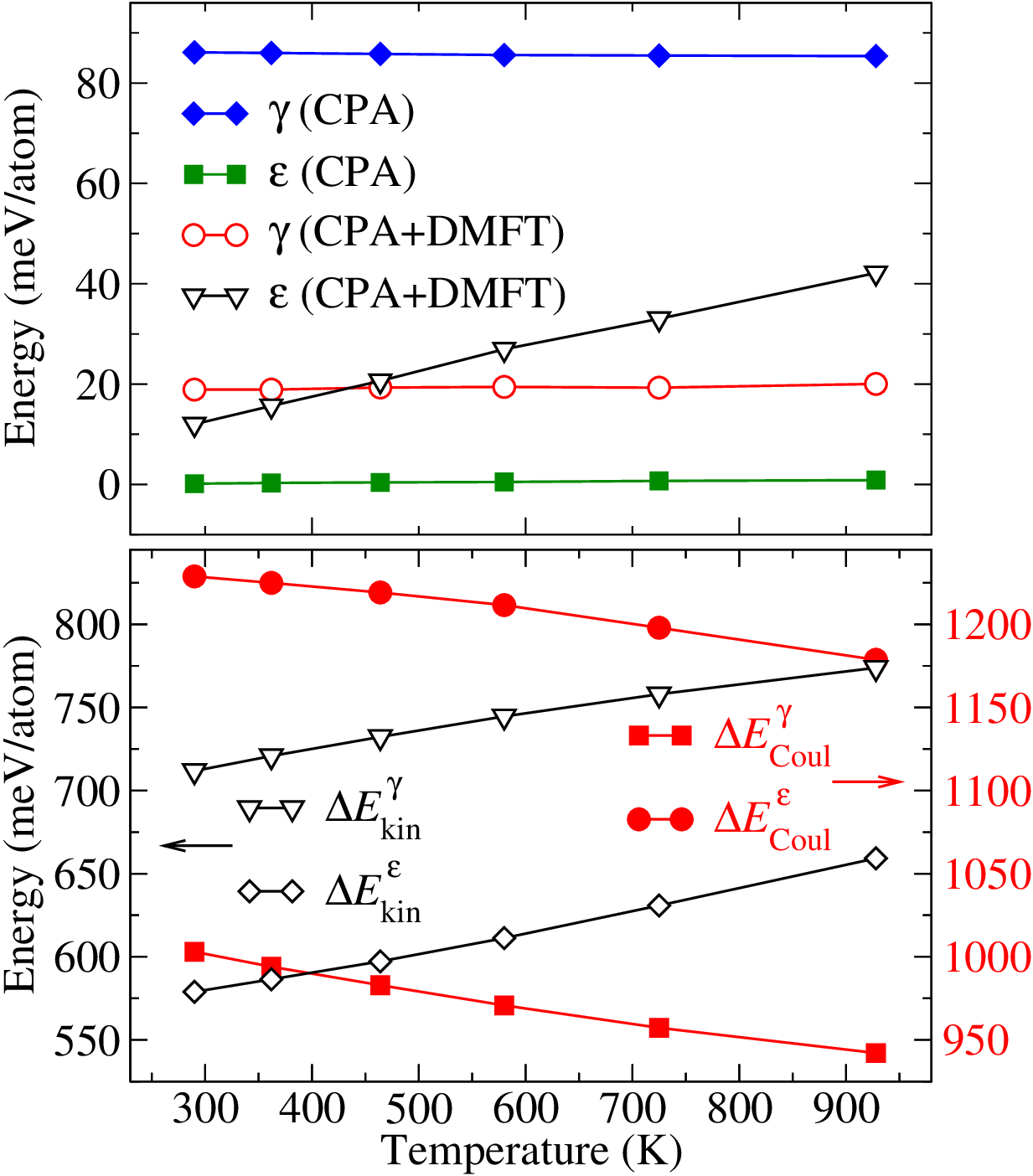}
\caption{ 
Total energy of the \g and \e phases of Fe$_{0.9}$Mn$_{0.1}$
obtained by CPA and CPA+DMFT (upper panel).
Kinetic and Coulomb contributions to total energy obtained by CPA+DMFT (lower panel).
Notations are the same as in section~\ref{sec:method}, namely, 
${\Delta E_{\rm Coul} = E_{\rm Coul}^{\rm CPA+DMFT} \!- E_{\rm DC}}$
and
${\Delta E_{\rm kin} = E_{\rm kin}^{\rm CPA+DMFT} \!- E_{\rm kin}^{0}}$.
The DFT total energies were shifted so that to obtain convenient values for total energies in the upper panel.
\label{fig:tot_energy}}
\end{figure}

In figure~\ref{fig:tot_energy} we present the obtained CPA and CPA+DMFT total energies
of Fe$_{0.9}$Mn$_{0.1}$ in the \g and \e phases.
The CPA alone resulted in a weak temperature dependence of total energies with
the \e phase being 86~meV/at lower in energy than the \g phase.
At the same time, our calculations by CPA+DMFT led to the stabilization of the \g phase at high temperatures.
In contrast to the \e phase, the total energy of \g phase
is almost temperature independent.
However, the kinetic and Coulomb contributions due to electronic correlations and disorder
have a strong temperature dependence in both phases (lower panel of figure~\ref{fig:tot_energy}).
In the case of \g phase, an increase of the kinetic contribution with temperature
is well compensated by the Coulomb contribution. 
%
%
The total energy curves of \g and \e phases intersect at ${T_0=440}$~K, 
which is close to the experimental \ge transition temperature of about 470~K~\cite{ts_experiment}.

The obtained results are weakly affected by the disorder parameters $V_s$ and $V_p$ for itinerant $s$ and $p$ states, respectively.
In particular, with ${V_s=V_p=0}$, the total energy curves intersect at 600~K for 10 at.~\% of Mn.
%
%
A simultaneous increase (decrease) of $V_d$ parameters for both phases for 0.05~eV 
results in a decrease (increase) of $T_0$ for about 100~K.
Since the energy difference between \g and \e phases is quite small,
the obtained results are sensitive to the Coulomb interaction parameters.
%
%
Calculations with ${U=3}$~eV resulted in ${T_0=1200}$~K,
while employing ${U=5}$~eV led to stabilization of the \g phase at all temperatures. 
However, the employed ${U=4}$~eV was calculated
using the $spd$ Wannier function basis set~\cite{Belozerov_u_paper}
and is in agreement with values from 3 to 4.5~eV 
obtained by cRPA calculations~\cite{cRPA}.

To assess the relative stability of phases, the Gibbs free energy ${G=E_{\rm total} + pV - TS}$ should be used instead of total energy. 
Since the \ge transition in Fe-Mn alloy is observed at the atmospheric pressure,
${p\,\Delta V^{\gamma-\varepsilon}}$ is only about ${10^{-4}}$~meV/at,
which is much smaller than the other contributions to the Gibbs energy difference between the phases.
%
Calculation of the entropy from first principles is still a challenging problem.
The entropy can be decomposed into the electronic, magnetic, vibrational and configurational contributions:
\begin{eqnarray} \label{entropy_total}
  S = S_{\rm el} + S_{\rm mag} + S_{\rm vib} + S_{\rm conf}.
\end{eqnarray}
%
%
The configurational entropy depends only on the concentrations of constituents.
Hence, the configurational entropy difference ${\Delta S_{\rm conf}^{\gamma-\varepsilon} = 0}$ for a given Mn content. 
The following simple estimates can be obtained for other entropy contributions. 

\begin{figure}[t!]
\centering
\includegraphics[clip=true, width=0.6\textwidth]{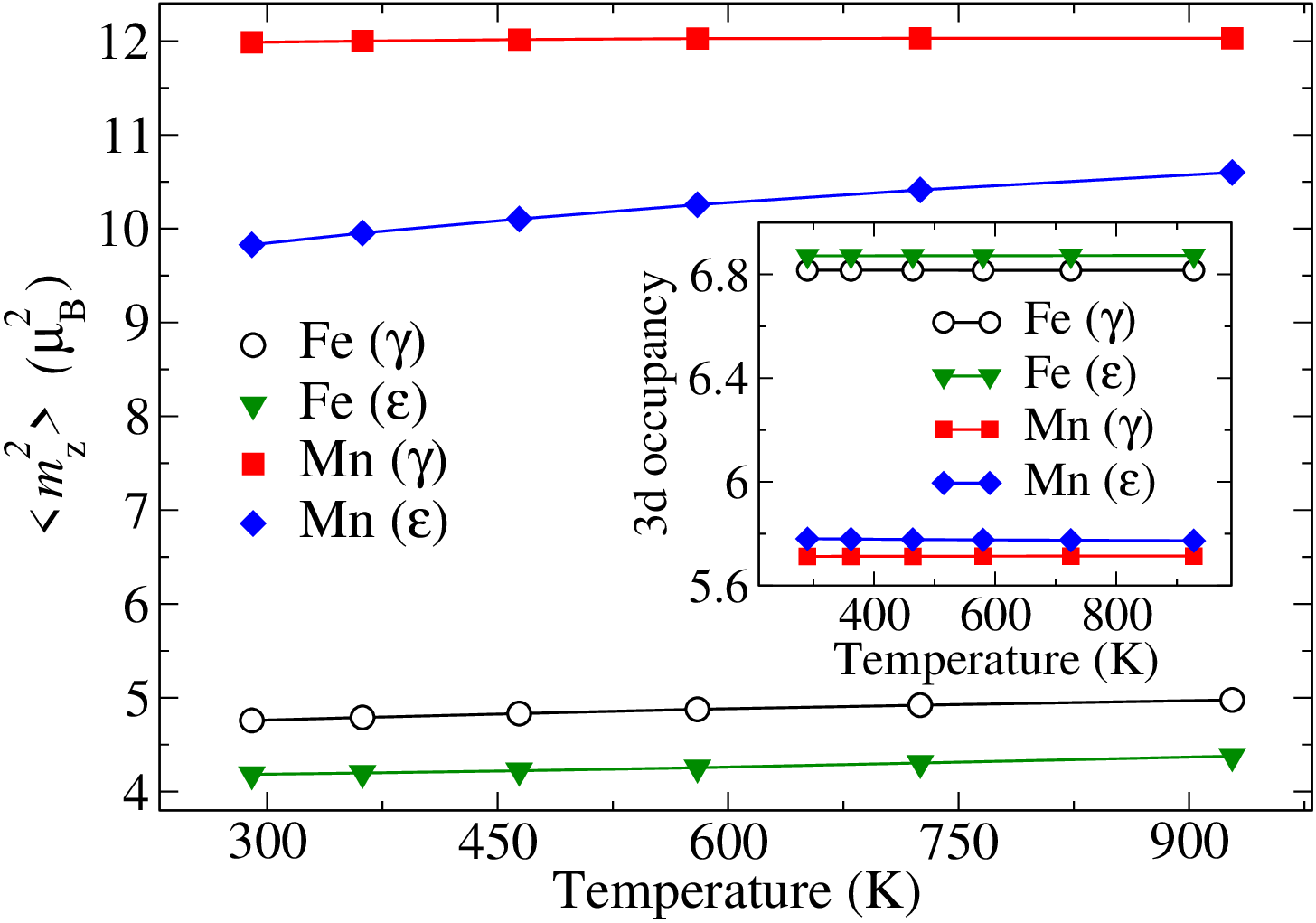}
\caption{ 
Average squared local magnetic moments of Fe and Mn atoms in Fe$_{0.9}$Mn$_{0.1}$ alloy
in \g and \e phases obtained by CPA+DMFT. Inset shows the occupancy of $3d$ states. 
\label{fig:mz}}
\end{figure}

The electronic entropy~\cite{electronic_entropy} can be expressed as
\begin{eqnarray} \label{entr_el}
  S_{\rm el} = &-& k_{\rm B} \int_{-\infty}^{+\infty} \{ f(\varepsilon)\, {\rm ln} f(\varepsilon) \nonumber \\
                &+& [1-f(\varepsilon)]\,{\rm ln}[1-f(\varepsilon)] \} N(\varepsilon)\, {\rm d}\varepsilon,
\end{eqnarray}
where $k_{\rm B}$ is the Boltzmann constant, $f(\varepsilon)$ is the Fermi function, $N(\varepsilon)$ is the density of states. 
The difference of electronic entropies ${\Delta S^{\gamma-\varepsilon}_{\rm el}}$
is almost temperature independent and is equal to ${-0.037~k_{\rm B}}$.
The magnetic entropy in the paramagnetic state can be expressed as
\begin{eqnarray} \label{magn_entr}
  S_{\rm mag} = k_{\rm B}\, {\rm ln}\hspace{0.01cm} (2m_{\rm av}+1).
\end{eqnarray}
Here, $m_{\rm av}$ is the average local magnetic moment,
which was calculated as a weighted sum of local moments on Fe and Mn atoms.
%
%
The local magnetic moment for constituent of type $i$ was estimated using  
average square of instantaneous local moment:
\begin{eqnarray}
\langle (m^i_{z})^2 \rangle =
\sum_{mm'\sigma}\left( \langle \hat{n}^i_{m\sigma} \hat{n}^i_{m'\sigma} \rangle 
- \langle \hat{n}^i_{m\sigma} \hat{n}^i_{m'\overline{\sigma}} \rangle  \right).
\end{eqnarray}
The obtained temperature dependence of squared local moments for 10 at.~\% of Mn
is presented in figure~\ref{fig:mz}.
%
%
%
The local moments in \g phase are found to be larger than those in \e phase. 
This is in agreement with results obtained by Reyes-Huamantinco \textit{et al.} for 22.5 at.~\% of Mn using DLM method~\cite{Reyes-Huamantinco2012}.
All local moments except those on Mn in \e phase have a weak dependence on temperature. 
The calculated difference of magnetic entropies ${\Delta S^{\gamma-\varepsilon}_{\rm mag}}$ is almost temperature independent 
and is equal to ${-0.057~k_{\rm B}}$.

The vibrational entropy difference can be expressed via the ratio of Debye temperatures as 
\begin{eqnarray} \label{vib_entr} 
  \Delta S^{\gamma-\varepsilon}_{\rm vib}
  = 3 k_{\rm B}\, {\rm ln}\, \frac{\Theta^{\gamma}_{\rm D}}{\Theta_{\rm D}^{\varepsilon}},
\end{eqnarray}
where $\Theta^{\gamma(\varepsilon)}_{\rm D}$ is the Debye temperature for $\gamma(\varepsilon)$~phase.
Using the approximation for Debye temperature derived by Moruzzi \textit{et al.}~\cite{Moruzzi1988} 
and experimental bulk moduli for \g and \e phases with 22.6 at.~\% of Mn~\cite{Lenkkeri_bulk_moduli},
we obtained ${\Theta^{\gamma}_{\rm D}/\Theta_{\rm D}^\varepsilon=1.052}$
resulting in ${\Delta S^{\gamma-\varepsilon}_{\rm vib} = 0.152~k_{\rm B}}$.
Taking into account all contributions, the total entropy difference ${\Delta S^{\gamma-\varepsilon} = 0.058~k_{\rm B}}$.
This value is close to 0.037~$k_{\rm B}$ obtained for the entropy change at the \ge
transition in pure Fe at 15 GPa using the Clausius-Clapeyron
equation and the slope of the phase boundary in pressure-temperature phase diagram.
Using the Gibbs free energies, we find the \ge transition at 530~K in good agreement with experiment.

\begin{figure}[t!]
\centering
 \includegraphics[clip=true, width=0.6\textwidth]{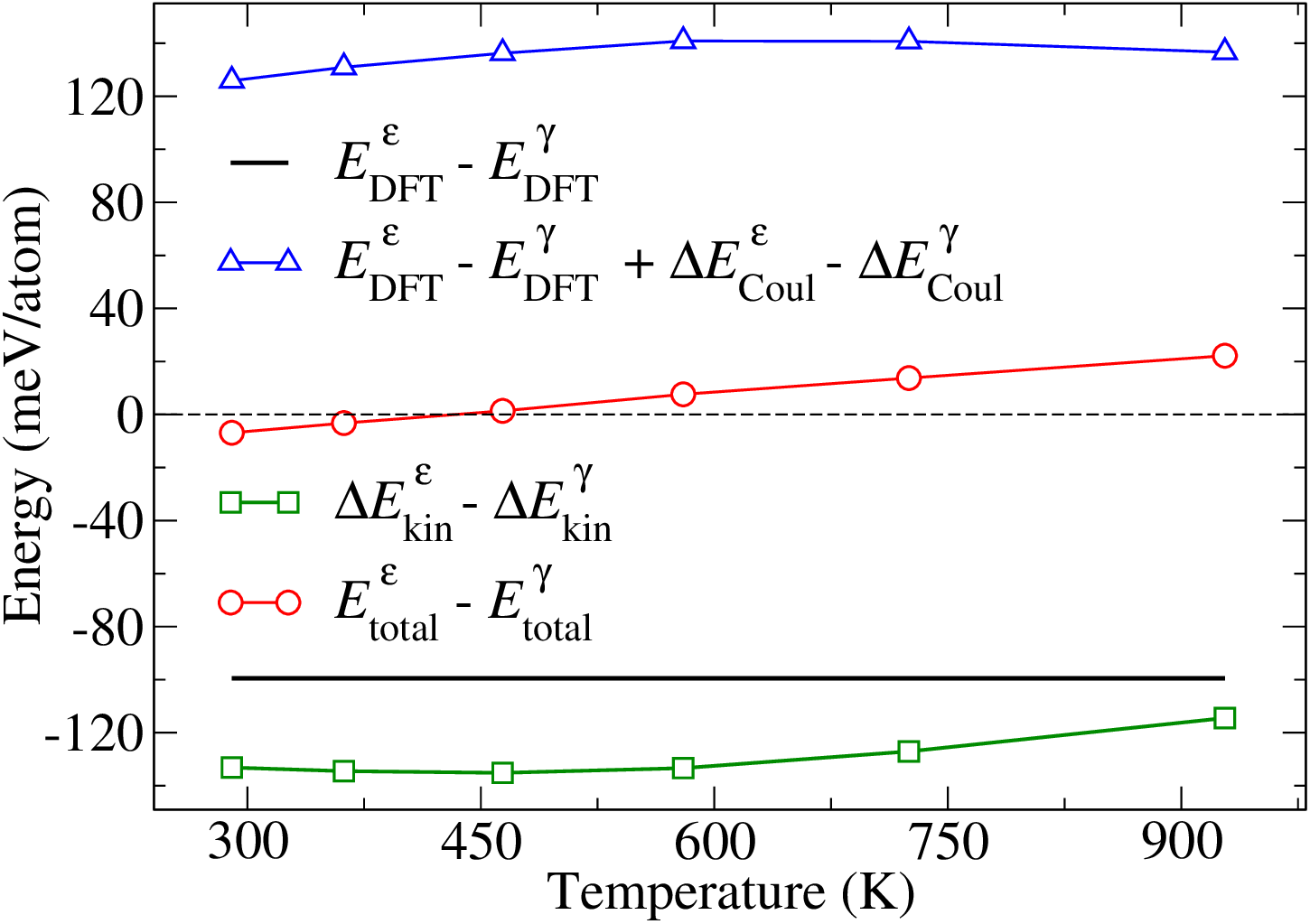}
\caption{ 
Energy difference between the \e and \g phases of Fe$_{0.9}$Mn$_{0.1}$ obtained by CPA+DMFT.
Notations are the same as in section~\ref{sec:method}, namely, 
${\Delta E_{\rm Coul} = E_{\rm Coul}^{\rm CPA+DMFT} \!- E_{\rm DC}}$
and
${\Delta E_{\rm kin} = E_{\rm kin}^{\rm CPA+DMFT} \!- E_{\rm kin}^{0}}$.
\label{fig:kin_and_coulomb}}
\end{figure}

To analyze the driving force behind the \ge transition,
in figure~\ref{fig:kin_and_coulomb} we present the contributions to energy difference between the \e and \g phases. 
%
%
%
We find that the temperature dependence of total energy difference is mainly from the Coulomb contribution
at low temperatures (${T<600}$~K) and from the kinetic contribution at high temperatures (${T>600}$~K).
At the same time, the last two terms in expression~(\ref{ekin_cpadmft}) for the kinetic energy 
make a negligible contribution, 
since the occupancies weakly depend on temperature in both phases (figure~\ref{fig:mz}).
Hence, the temperature dependence of the kinetic energy difference is mainly due to the first term in expression~(\ref{ekin_cpadmft}),
which includes both the electronic correlations and disorder effects. 
%
%
%
%

The magnetic correlation contribution to the Coulomb energy can be approximately expressed as
${E_{\rm Coul}^{ \rm magn}=-\frac{1}{4}I\langle m_z^2 \rangle}$, where ${I=\frac{1}{5}(U+4J)}$.
Squared local moments on Fe atoms have similar temperature behaviour in both phases (figure~\ref{fig:mz}),
and their difference slightly decreases from 0.6 to 0.57~$\mu_{\rm B}^2$ upon cooling from 930 to 290~K,
lowering the energy of \e phase with respect to \g phase.
The opposite behaviour is observed for local moment on Mn, which decreases faster upon cooling in \e phase than in \g phase,
favouring the stabilization of \g phase. 
However, one should keep in mind that the Mn contribution is significantly suppressed at given concentrations.
The obtained results indicate that both the Coulomb and kinetic contributions play an important role at the \ge transition in Fe-Mn alloys. 
This is in contrast to the \ag transition in pure iron where 
the magnetic correlation energy was shown to be an essential driving force behind this transition~\cite{Leonov_bccfcc}.

%

\begin{figure}[t!]
\centering
\includegraphics[clip=true, width=0.55\textwidth]{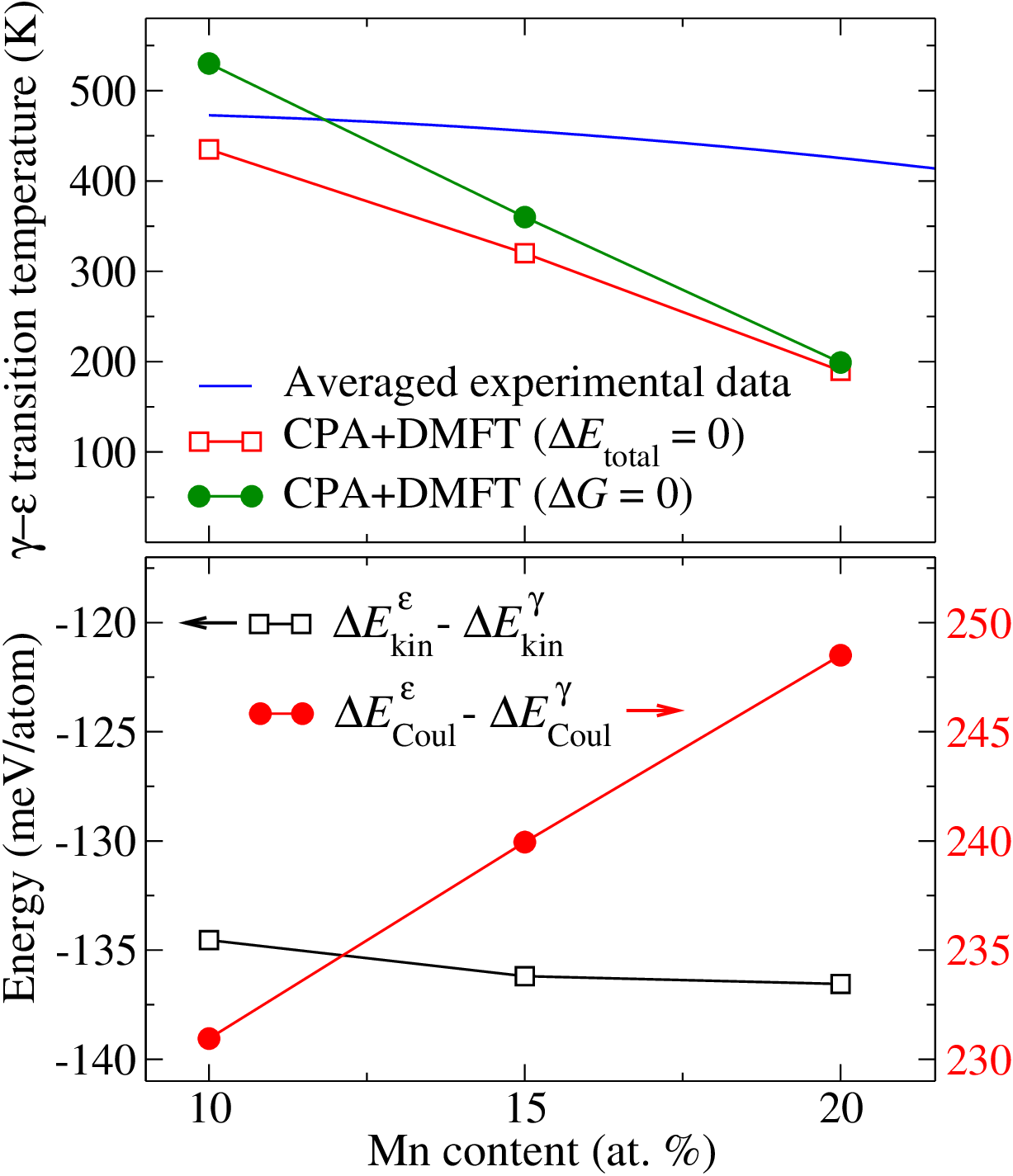}
\caption{ 
Temperatures of \ge transition (upper panel) calculated by CPA+DMFT in comparison with 
averaged experimental data~\cite{Lee}.
Energy difference between the \e and \g phases (lower panel) obtained by CPA+DMFT at 360~K.
Notations are the same as in section~\ref{sec:method}, 
${\Delta E_{\rm Coul} = E_{\rm Coul}^{\rm CPA+DMFT} \!- E_{\rm DC}}$,
${\Delta E_{\rm kin} = E_{\rm kin}^{\rm CPA+DMFT} \!- E_{\rm kin}^{0}}$.
\label{fig:ts}}
\end{figure}

In figure~\ref{fig:ts} we present the \ge transition temperature
as a function of Mn concentration. 
%
%
The total entropy difference weakly depends on temperature and is equal to ${0.067~k_{\rm B}}$ for 20~at.~\% of Mn.
The calculated transition temperature decreases with increasing Mn content from 10 to 20 at.~\%
in agreement with the experimental data~\cite{Lee}.
However, the difference between the calculated and experimental transition temperatures grows with Mn content.
This difference is less than 100~K for 10 and 15 at.~\% of Mn, while it is about 230~K for 20 at.~\% of Mn.
This can be caused by the fact that the effective medium approach employed in CPA and DMFT gives better results at low concentrations.
To identify the specific role of Mn, in the lower panel of figure~\ref{fig:ts}
we present the difference of Coulomb and kinetic energies between the phases at 360~K.
We find that the Coulomb energy is responsible for the decrease of the \ge transition temperature.
%
%
This can be explained by increasing contribution of Mn to the Coulomb energy of the alloy,
while the magnetic correlation energy of Mn favours the stabilization of \g phase at low temperatures.

\section{Conclusions\label{sec:conclusions}}

We presented a computational scheme for total energy calculations of disordered alloys with strong electronic correlations.
It employs the CPA+DMFT approach treating electronic correlations and disorder on the same footing.
%
%
The proposed computational scheme can be used to study
correlation-induced structural and/or magnetic transitions as well as related properties 
in paramagnetic and magnetically ordered phases of disordered systems.
In particular, we applied it to study the \ge structural transition in paramagnetic Fe$_{1-x}$Mn$_{x}$ alloys with $x$ from 0.1 to 0.2.
The calculated transition temperature is in a good agreement with the experimental data.
The local magnetic moment on Mn was found to have more pronounced temperature dependence in \e phase than in \g phase.
Upon cooling, this leads to the lowering the energy of \g phase with respect to \e phase due to magnetic correlation energy.
Both the Coulomb and kinetic energies were demonstrated to contribute to the \ge transition. 
This is in contrast to the \ag transition in pure Fe,
where the magnetic correlation energy alone was shown
to be responsible for the structural transformation~\cite{Leonov_bccfcc}.
However, one should keep in mind that the kinetic energy in CPA+DMFT approach includes both the electronic correlations and disorder effects which cannot be separated. 
%
%
%
%
%

Considering the alloys with Mn content from 10 to 20~at.~\%,
we found that the decrease of the \ge transition temperature is caused by the Coulomb energy.
This agrees well with the above mentioned finding that
the magnetic correlation energy of Mn favours the stabilization of \g phase at low temperatures. 
The obtained results indicate that the CPA+DMFT approach is a promising tool for studying
the real substitutional alloys with strong electronic correlations.

%
%
%
%
%

\section*{Acknowledgments}
 
The authors are grateful to M A Korotin and Yu N Gornostyrev for useful discussions.
The study was supported by the grant of the Russian Scientific Foundation (project no. 14-22-00004).

\end{document}